\documentclass[twocolumn,showpacs]{revtex4}
\usepackage[latin9]{inputenc}
\usepackage{amsmath}
\usepackage{amssymb}
\usepackage{esint}

\makeatletter
\@ifundefined{textcolor}{}
{%
 \definecolor{BLACK}{gray}{0}
 \definecolor{WHITE}{gray}{1}
 \definecolor{RED}{rgb}{1,0,0}
 \definecolor{GREEN}{rgb}{0,1,0}
 \definecolor{BLUE}{rgb}{0,0,1}
 \definecolor{CYAN}{cmyk}{1,0,0,0}
 \definecolor{MAGENTA}{cmyk}{0,1,0,0}
 \definecolor{YELLOW}{cmyk}{0,0,1,0}
}

\def\be{\begin{eqnarray} &&}

\def\ee{\end{eqnarray}}

\def\bew{\begin{widetext}}
\def\ew{\end{widetext}}

\makeatother

\begin{document}

\title{A comment on chiral restoration at finite baryon density in hyperspherical
unit cells}

\author{Hilmar Forkel}

\affiliation{Institut für Physik, Humboldt-Universität zu Berlin, D-12489 Berlin,
Germany}
\begin{abstract}
Prompted by recent work of Adhikari, Cohen, Ayyagari and Strother
\emph{On chiral symmetry restoration at finite density in large-$N_{c}$
QCD} (Phys. Rev. C \textbf{83}, 065201 (2011)), we revisit the description
of dense baryonic matter in terms of hyperspherical unit cells. We
focus mainly on the interpretation of the unique energy, curvature
and symmetry properties which enable such $S^{3}$ cells to describe
full chiral restoration in Skyrme models and which markedly distinguish
them from the flat and periodic unit cells of Skyrmion crystals. These
key features clarify, in particular, why an $S^{3}$ cell interpretation
as a crystal-cell model in which the specific cell geometry is without
physical significance, as tentatively adopted by Adhikari et al.,
is insufficient. The ensuing criticism does therefore not apply to
the usual interpretation of $S^{3}$ cells which we describe. We also
suggest a few directions in which the latter interpretation may be
developed further.
\end{abstract}

\pacs{11.15.Pg, 12.38.Aw, 21.65.-f}

\keywords{Chiral symmetry restoration}

\maketitle

\section{Introduction and motivation\label{sec:intro}}

The recent work of Adhikari et al. on chiral symmetry restoration
at high baryon densities in QCD with a large number $N_{c}$ of colors
\cite{adh11} contains a section which discusses chiral restoration
in the ``hypersphere approach\textquotedblright{} \cite{man86,man87,for89}.
The latter describes dense matter in models of Skyrme type \cite{zah86}
by means of generalized, curved unit cells with the geometry of a
three-dimensional sphere $S^{3}$. This geometry was selected because
the interplay of its symmetry group SO$\left(4\right)$ with the chiral
SU$\left(2\right)\times$ SU$\left(2\right)\simeq$ SO$\left(4\right)$
symmetry of the dynamics uniquely enables such hyperspherical cells
to model chiral restoration beyond a critical density \cite{man86,man87}.
In fact, $S^{3}$ is the \emph{only} unit-cell geometry in which the
Skyrmion can attain its absolute energy minimum and in which the transition
to a chirally restored phase can take place.

The above properties provided the original motivation for studying
$S^{3}$ cells in the Skyrme model \cite{man86,man87}. At this stage
no analogies with the flat unit cells of periodic Skyrmion arrays
were made. Only later it was realized that the discrete ``half-Skyrmion''
symmetry \cite{halfSkyrm}, which emerges in the high-density phase
of Skyrmion crystals \cite{array}, is restored together with the
chiral group on $S^{3}$ as well \cite{for89}. Since several consequences
of full chiral restoration on $S^{3}$ turned out to require just
the half-Skyrmion symmetry, it was further argued in Ref. \cite{for89}
that the emergence of the latter should be interpreted as signaling
chiral restoration in the restricted setting of Skyrmion crystals.
Moreover, calculations in $S^{3}$ cells proved to be far less complex
than those in flat, periodic unit cells %
\footnote{Obviously, computational simplicity is a welcome side benefit which
makes analytical studies possible, in contrast to crystal calculations
(beyond the Wigner-Seitz approximation) which require numerical computations.
Nevertheless, this is not a major reason for preferring $S^{3}$ (although
this impression has sometimes been given in the literature). In fact,
computations on several other, similarly symmetric compact manifolds
are not more involved (cf. Refs. \cite{man87} and \cite{bra08} for
explicit examples).%
}. The crucial impact of the specific hypersphere properties, especially
on the density-dependent multiplet structure of the Goldstone bosons
and other excitations, furthermore strengthened the original view
that $S^{3}$ cells provide an independent and with regard to chiral
symmetry properties more complete description of dense matter.

A rather different interpretation of hyperspherical cells was recently
explored in Ref. \cite{adh11}. Motivated by the suggested quarkyonic
dense-matter phase at large $N_{c}$ \cite{mcl07}, the bulk of Ref.
\cite{adh11} investigates chiral restoration conditions in Skyrme
models and in large-$N_{c}$ QCD. This raised the question to what
extent $S^{3}$ cells can be considered as faithful models for the
flat, periodic unit cells of Skyrmion crystals. To address this issue,
the authors of Ref. \cite{adh11} adopt the position that the sole
purpose of hyperspherical unit cells should be to ``approximate a
Skyrmion in the crystal'' %
\footnote{A similar view may have occasionally been implied in the earlier literature
on the subject.%
}. In addition, they assume ``that the principal effect of putting
a Skyrmion into a crystal is to restrict the space over which it can
spread'' and ``that\textbf{ }using a hypersphere to restrict the
volume of the Skyrmion acts generically like other restrictions on
its volume''. Below we will explain why these assumptions, which
would deprive the cell geometry of its physical significance \cite{adh11},
are oversimplifications. Indeed, they ignore the unique energy, symmetry
and curvature properties of the $S^{3}$ geometry and would not even
hold for the flat unit cells of Skyrmion crystals. In addition, these
premises lead the authors of Ref. \cite{adh11} to the unduly pessimistic
conclusions that ``the special properties of the geometry\textbf{
}...\textbf{ }make the (gained) intuition totally unreliable even
for qualitative issues associated with chiral symmetry breaking and
its possible restoration in the average sense'' and that the ``evidence
for chiral restoration ... was an artifact of the hyperspherical geometry''.

The main purpose of the present note is to clarify that the above
criticism, based on the problematic interpretation of Ref. \cite{adh11},
does not apply to the standard $S^{3}$ cell interpretation. To this
end, we will discuss the physical significance of the $S^{3}$ geometry
\footnote{This may be of some independent use because the foundations of the
hypersphere approach were occasionally simplified in the literature. %
}, address the problems with the premises underlying the interpretation
of Ref. \cite{adh11}, and point out an important difference between
chiral restoration ``in the average sense'' in flat space and curved
cells. Finally, we will suggest a possible extension of the $S^{3}$
unit cell interpretation.

\section{The unique ``chiral\textquotedblright{} significance of $S^{3}$
unit cells}

\label{sec:unqrole}

We start by reviewing those key features of hyperspherical unit cells
which first suggested their physical significance. These properties
will also help to explain what prevents the related results from being
``an artifact of the (unphysical) choice of geometry'', as they
would appear to be in the problematic interpretation of Ref. \cite{adh11}.
In fact, in models of Skyrme type \cite{zah86} several essential
dense-matter properties turned out to be uniquely encoded into $S^{3}$
unit cells \cite{man86,man87}. Manton demonstrated this uniqueness
by considering generalized Skyrmions as topologically nontrivial maps
between two Riemannian manifolds, i.e. $\Sigma_{\text{cell}}$ (the
unit cell space) and $\Sigma$ (the target space in which the fields
take values) \cite{man87}. He noted, in particular, that ``the metrics
on both $\Sigma_{\text{cell}}$ and $\Sigma$ are \emph{essential}\textquotedblright{}
and that ``the energy of the Skyrmion is a measure of the geometrical
distortion induced by the map\textquotedblright{}. In fact, since
chiral symmetry allows only gradient interactions among the pions
and since those are exceptionally sensitive to the local background
curvature, one expects an enhanced impact of the $\Sigma_{\text{cell}}$
metric on even qualitative dense-matter predictions. 

As in the nonlinear $\sigma$ model, the chiral symmetry of the Skyrme
dynamics is nonlinearly realized \cite{col69} on its unbroken isospin
subgroup SU$\left(2\right)\sim S^{3}$, i.e. $\Sigma=S^{3}$. (For
simplicity, we assume exact chiral SU$\left(2\right)\times$ SU$\left(2\right)$
symmetry of the dynamics and exact SU$\left(2\right)$ isospin symmetry
of the vacuum.) Hence the energetically privileged role of $\Sigma_{\text{cell}}=S^{3}$
emerges already at this qualitative level. More quantitatively, the
interplay between the Skyrmion's energy and topology results in an
absolute energy minimum given by the Bogomol'ny (or Faddeev) bound
\cite{zah86}. Since the Skyrmion's energy functional is an efficient
measure of the metric deformation between the unit cell $\Sigma_{\text{cell}}$
and the target space $\Sigma$, the field configuration which saturates
the Bogomol'ny bound should not induce any such deformation. This
requires both $\Sigma_{\text{cell}}$ and $\Sigma$ to have the same
metric. Hence $S^{3}$ is the unique unit-cell geometry in which the
Skyrmion can become the metric preserving identity map %
\footnote{We recall that the identity map between $S^{3}$'s of \emph{different}
radii is not always the map of lowest energy. In fact, the identity
map is stable only for radii $L$ of $\Sigma_{\text{cell}}=S^{3}\left(L\right)$
smaller than the critical radius at which the bound is saturated and
the chiral restoration transition takes place \cite{man87}. For larger
radii, in contrast, the Skyrmion starts to localize on $S^{3}$ as
on $\text{\ensuremath{\mathbb{R}}}^{3}$.%
} and thereby attain its absolute energy minimum.

In addition, $S^{3}$ is the unique space which SO$\left(4\right)$
transformations leave invariant. The Skyrmion's ``hedgehog'' coupling
between space and isopace links this isometry group to the chiral
SU$\left(2\right)\times$ SU$\left(2\right)\simeq$ SO$\left(4\right)$
group which acts analogously on the internal field space $\Sigma=S^{3}$.
This provides the key to understanding why chiral restoration (beyond
the critical density and in a sense to be specified below) is possible
only in $S^{3}$ unit cells. Indeed, the symmetry group $G$ of the
Hamiltonian (for static fields) in general unit cells $\Sigma_{\text{cell}}$
is the product of the spatial cell symmetries $G_{\text{cell}}$ and
of the chiral group $\text{SO}\left(4\right)_{\chi}\simeq\text{SO}\left(3\right)_{L}\times\text{SO}\left(3\right)_{R}$.
In the presence of a semi-classically quantized Skyrmion, the symmetry
of the spectrum is reduced to the subgroup of $G$ which leaves the
Skyrmion invariant. Due to the hedgehog-type coupling $\hat{x}^{i}\tau^{i}$
in the Skyrmion solution $U\left(\vec{x}\right)=\exp\left[i\hat{x}^{i}\tau^{i}F\left(\left|\vec{x}\right|\right)\right]$
this group includes the diagonal subgroup of spatial and SO$\left(3\right)_{\text{iso}}$
isospin rotations (where the latter belong to the diagonal subgroup
of SO$\left(4\right)_{\chi}$).

In the familiar flat-space example $\Sigma_{\text{cell}}=\mathbb{R}^{3}$
one thus has $G_{\text{cell}}=T\left(\mathbb{R}^{3}\right)\times$
SO$\left(3\right)_{\text{rot}}$ where $T\left(\mathbb{R}^{3}\right)$
are the translations in $\mathbb{R}^{3}$. Since the localized, classical
Skyrmion breaks translational invariance while isospin rotations can
be undone by spatial rotations around the Skyrmion's center, only
the diagonal subgroup $\text{SO}(3)_{\text{grand}}=\text{diag}\left\{ \text{SO}\left(3\right)_{\text{rot}}\times\text{SO}(3)_{\text{iso}}\right\} $
consisting of simultaneous spatial and isospin rotations leaves the
Skyrmion invariant. In other words, chiral (as well as rotational
and translational) symmetry is spontaneously broken to the so-called
``grand spin'' from which isospin (and rotational) symmetry can
be recovered by projection \cite{adk83}.

We now return to the description of dense matter and search for a
three-dimensional unit cell $\Sigma_{\text{cell}}$ whose symmetry
enables a full chiral $\text{SO}\left(4\right)$ group to leave the
Skyrmion invariant. The hedgehog coupling suggests to write $G{}_{\text{cell}}=T\left(\Sigma_{\text{cell}}\right)\times\text{SO}(3)_{\text{rot}}$
as the product of the rotations around the Skyrmion center and the
coset of ``generalized translations\textquotedblright{} $T\left(\Sigma_{\text{cell}}\right):=G_{\text{cell}}/\text{SO}(3)_{\text{rot}}$
which move this center around the cell. Since $\text{SO}(3)_{\text{rot}}$
is linked to $\text{SO}(3)_{\text{iso}}$ as above, with the Skyrmion
leaving only their diagonal subgroup invariant, we look for an extension
of the grand-spin subgroup to $\text{SO}\left(4\right)$. This requires
the full $\text{SO}\left(4\right)_{\chi}$ to take part in one factor,
and the latter to be multiplied by the extension of $\text{SO}(3)_{\text{rot}}$
to an analogous $G{}_{\text{cell}}=SO\left(4\right)$. The unique
cell with this symmetry group is $\Sigma_{\text{cell}}=S^{3}$. Hence
the hypersphere is indeed the only cell geometry in which the invariance
group of the Skyrmion can become $\text{SO}\left(4\right)$. 

Two generic situations must now be distinguished. For cell radii $L$
which are large compared to the Skyrmion's size, the Skyrmion is localized
on $S^{3}$ and therefore $T\left(S^{3}\right)$ is spontaneously
broken. For $L$ equal to or smaller than a critical radius, on the
other hand, the Skyrmion on $S^{3}$ delocalizes completely. As discussed
above, at the critical radius it becomes the energy-minimizing identity
map between $\Sigma_{\text{cell}}=S^{3}$ and $\Sigma=S^{3}$. Hence
no center is singled out anymore either in $\Sigma_{\text{cell}}$
or in $\Sigma$, and any generalized translation can be compensated
by the corresponding ``translation\textquotedblright{} from the axial
coset $\text{SO}\left(4\right)_{\chi}/\text{SO}(3)_{\text{iso}}$
on the field manifold. The symmetry group of the identity-map Skyrmion
is therefore $\text{SO}\left(4\right)_{\chi^{\prime}}=\text{diag}\left\{ \text{SO}\left(4\right)_{\text{cell}}\times\text{SO}\left(4\right)_{\chi}\right\} $.
After projection as above, this $\text{SO}\left(4\right)_{\chi^{\prime}}$
turns into the standard chiral group which is thus indeed restored.

The above, complete chiral restoration implies that all averaged chiral
order parameters disappear\textbf{,} as noticed in Ref. \cite{adh11},
and leaves crucial imprints on the fluctuation spectrum around the
Skyrmion in the $S^{3}$ cell \cite{for89}. Below the critical density
the excitations fall (after projection) into isospin multiplets and
include a triplet of massless Goldstone pions, i.e. the telltale signature
of spontaneously broken chiral symmetry. At and beyond the critical
density, on the other hand, the spectrum is classified by the larger,
chiral $\text{SO}\left(4\right)$ group. The former Goldstone bosons,
in particular, join their three parity partners in a degenerate chiral
multiplet whose mass increases with the density \cite{for89}, as
expected from complete chiral restoration. 

Several additional features of the above restoration mechanism were
studied later and revealed, for example, an interesting interplay
with kaon condensation \cite{for90}. For more recent work on Skyrmions
in hyperspherical cells see Ref. \cite{moreS3}. An interesting appearance
of $S^{3}$ cells in the context of holographic QCD \cite{kim08}
is related to instantons on $S^{3}$ \cite{for04} which generate
approximate Skyrmion solutions by means of the Atiyah-Manton map \cite{sam89}.

\section{Additional aspects of the $S^{3}$ cell interpretation \label{sec:more}}

The key properties discussed above provide the basis for the physical
interpretation of $S^{3}$ Skyrmion cells. Nevertheless, some of their
more unconventional features, including several of those which set
them even qualitatively apart from flat, periodic unit cells, still
await a better understanding. In the present section we suggest a
few directions in which the understanding of $S^{3}$ cells may potentially
be improved, and we address the problems with the interpretation of
Ref. \cite{adh11} in more detail.

We start by recalling that $S^{3}$ unit cells were found to describe
principal features of dense matter even in chiral models which are
\emph{not} of Skyrme type. In the remarkably different Nambu Jona-Lasinio
(NJL) model \cite{nam61}, in particular, where chiral symmetry is
broken by interactions among quarks which carry intrinsic baryon number
but no topology, a growing baryon density described by $S^{3}$ unit
cells was shown to trigger the transition to the chirally restored
phase as well \cite{for92}. Simultaneously, at a critical density
consistent with standard values, the previously massless and tightly
bound Goldstone pions disappear \cite{for95}. Again the increasing
curvature of the $S^{3}$ unit cells, and not just their reduced volume,
was found to play an explicit dynamical role in achieving chiral restoration.
(In addition, the $S^{3}$ description naturally generalizes to finite
temperature and reveals interesting analogies between the ``geometric''
implementations of temperature and density \cite{for295}.) The above
results suggest that the $S^{3}$ unit-cell geometry encodes \emph{chiral}
interactions of nucleons with the ambient matter \cite{for95}, in
addition to those encoded in the flat-space model Lagrangian %
\footnote{Problems encountered with MIT bags in the $S^{3}$ unit cell \cite{wei92}
may therefore be related to the chiral-symmetry breaking bag boundary.%
}. (The distortions of the $S^{3}$ geometry considered in Ref. \cite{adh11}
may thus be regarded as admixing additional interactions with the
ambient matter, described by the pions' interactions with the deformed
cell background. The latter break chiral symmetry explicitly and thus
prevent exact chiral restoration.)

When attempting to put the above interpretation of the cell curvature
as mediating chiral interactions with the surrounding baryons on a
more solid basis, the lower bound on the Skyrmion energy, which can
be saturated only on $S^{3}$, provides a valuable hint. Indeed, one
may in principle determine the cell geometry at a given density variationally,
as done e.g. in condensed-matter physics. When minimizing the cell
energy, including the contributions from interactions with the surroundings,
one may then allow by some stretch of the imagination not just the
flat, extrinsic geometry (i.e. distances and boundary conditions)
but even the intrinsic curvature of the cell to vary \cite{for95}.
According to the arguments of Sec. \ref{sec:unqrole} the resulting
cell geometry should then be $S^{3}$, at least beyond the critical
density. In analogy to translating interactions with surrounding baryons
into the variationally determined structure of flat unit cells, the
$S^{3}$ curvature would then indeed encode additional, chiral interactions
of the Skyrmion with the ambient matter.

Even with such a potential dynamical origin of $S^{3}$ unit cells
in mind, however, it still seems counterintuitive that their instrinsic
curvature and missing boundary prevent them from being embedded into
flat space. Nevertheless, such curved cells should represent identical
units of a self-repeating structure which describes a finite average
baryon density, to be identified with the inverse of their volume,
over macroscopic distances. In order to guess a potential explanation
for the above observations it seems again helpful to draw intuition
from the unique capability of $S^{3}$ cells to restore chiral symmetry.
The latter agrees with QCD expectations and goes far beyond the discrete
half-Skyrmion (sub-) symmetry which Skyrmion arrays can restore %
\footnote{Higher-dimensional order parameters \cite{for89,adh11} and a dense-matter
generalization of the pion decay constant \cite{lee03} do not vanish,
for example, when only the half-Skyrmion symmetry is restored. (A
more genuine chiral restoration was proposed to occur in Skyrmion
arrays when the expectation value of an additional dilaton field vanishes
\cite{lee203}.)%
}. Hence $S^{3}$ cells represent at least this crucial aspect of chiral
dense-matter physics more completely than flat, periodic unit cells,
and this is possible because they in a sense (cf. Sec. \ref{sec:unqrole})
restore the translational symmetry which crystals break spontaneously. 

Continuing this line of thought, it is tempting to speculate that
$S^{3}$ cells cannot be embedded into $\mathbb{R}^{3}$ and directly
match on to ``adjacent'' cells because -- unlike flat, periodic
cells -- they do not just encode interactions with their immediate
``neighbors''. More specifically, their intrinsic curvature may
contain information on the averaged interactions with more distant
or even all other cells. The non-locality of such averages could then
reflect itself in a ``dissolution'' of the cell boundary. Moreover,
the implied averaging procedure should be able to restore translational
symmetry, which may be a prerequisite for full chiral restoration.

Rather than contemplating additional physics which may potentially
be encoded in the cell curvature, the minimalistic interpretation
of Ref. \cite{adh11} takes the opposite route. It tentatively ignores
even the impact of the extrinsic cell geometry and postulates that
at least the qualitative physics should not depend on it (cf. Sec.
\ref{sec:intro}). Hence the role of the cell is reduced to just providing
a computationally convenient volume with essentially arbitrary geometry
to constrain the spreading of the Skyrmion. In view of the $S^{3}$
geometry's unique impact discussed in Sec. \ref{sec:unqrole}, this
assumption cannot be even qualitatively correct. Instead, the cell
geometry (and topology) matters even at the qualitative level, as
it does in the flat unit cells of conventional Skyrmion crystals with
their remarkable sensitivity to the boundary conditions. (In fact,
without the latter the half-Skyrmion symmetry would not emerge.) On
$S^{3}$ the impact of the geometry is further enhanced by the heightened
sensitivity of the chiral dynamics to the background curvature %
\footnote{As a case in point, several different geometries were found to generate
qualitatively different physics in the same cell volume \cite{man87,bra08}.%
}. (Since the unit cell's boundary conditions determine the crystal
structure up to scales, their neglect would be inadequate in condensed-matter
physics as well, incidentally.)

Finally, it is instructive to reflect upon the impact of Ref. \cite{adh11}'s
main result on the interpretation of $S^{3}$ cells. The authors of
Ref. \cite{adh11} argue that all spatially-averaged chiral order
parameters in flat-space Skyrme models and large-$N_{c}$ QCD can
simultaneously vanish only if chiral symmetry is also restored in
the conventional, local sense (as signaled by the vanishing quark
condensate in QCD). Since at least naively the former seems to happen
without the latter in $S^{3}$ cells, one may suspect a contradiction
with the hypersphere description of dense matter %
\footnote{One could try to evade such a contradiction from the outset, of course,
by tentatively assuming that $S^{3}$ cells incorporate relevant $1/N_{c}$
corrections and thereby generate a phase diagram which is closer to
QCD (with $N_{c}=3$) expectations.%
}. As a matter of fact, the problematic interpretation of Ref. \cite{adh11}
creates such a contradiction by postulating that $S^{3}$ cells should
faithfully model the flat unit cells of Skyrmion crystals. 

In the interpretation of $S^{3}$ cells as an independent description,
on the other hand, it may at first appear that averting a contradiction
requires some of the physics encoded in hyperspherical cells to differ
from that of dense matter in Skyrme models and large-$N_{c}$ QCD.
Given the simplicity of the $S^{3}$ cell description and the lack
of a first-principles derivation, this conclusion would not even be
surprising. It is important to realize, however, that it would also
be premature. This is because spatial averaging over the $S^{3}$
cell does\emph{ }not\emph{ }have to translate into uniform spatial
averaging over some flat-space configurations, including those which
the $S^{3}$ cells are supposed to describe %
\footnote{In fact, even in the (not viable) interpretation of $S^{3}$ cells
as faithful models of flat unit cells it would not be immediately
clear whether and how a pointwise correspondence between the cell
volumes could be established.%
}. In the tentative interpretation of the previous paragraphs this
becomes particularly obvious because in such a scenario the volume
of a particular $S^{3}$ cell does not even correspond to a specific
and exclusive volume of dense matter in flat space. 

Hence one should keep in mind that chiral restoration ``in the \emph{flat-space}
average sense'' as considered in Ref. \cite{adh11} is not identical
to chiral restoration ``in the curved-cell average'' as it occurs
in $S^{3}$ cells. As a consequence, there is \emph{a priori} no reason
for conclusions regarding chiral restoration in the flat-space average
sense, including those of Ref. \cite{adh11}, to apply to chiral restoration
in $S^{3}$ cells as well. In fact, as alluded to above one may optimistically
hope that the latter describes a situation which corresponds more
closely to the conventional, i.e. local chiral restoration in dense
matter than to the flat-space averaged version dealt with in Ref.
\cite{adh11}.

\section{Conclusion}

\label{sec:concl}

In this note we clarify that the interpretation of the hypersphere
approach to dense matter as tentatively considered in Ref. \cite{adh11}
is based on inadequate premises. Although not always obvious in its
presentation, the criticism of Ref. \cite{adh11} does therefore not
apply to the standard interpretation and its chiral-restoration mechanism
as described above. In particular, the standard interpretation does
not require hyperspherical cells to be models of flat unit cells (several
analogies and shared features notwithstanding), and it ascribes specific
physical significance to the cell geometry as encoding chiral interactions
with the ambient matter. We furthermore point out differences between
chiral restoration ``in the spatial-average sense'' in flat space
and in curved cells, and we provide a few additional suggestions concerning
the dynamical origin and potential physics content of $S^{3}$ cells.

We thank Tom Cohen for extensive correspondence on the interpretation
of the $S^{3}$ approach adopted in Ref. \cite{adh11}.

\end{document}